# Route-cost-assignment with joint user and operator behavior as a many-to-one stable matching assignment game

**Saeid Rasulkhani, Joseph Y. J. Chow***

C$^2$SMART University Transportation Center, Department of Civil & Urban Engineering, New York University, New York, NY, USA

*Corresponding author email: joseph.chow@nyu.edu

**Abstract**
We propose a generalized market equilibrium model using assignment game criteria for evaluating transportation systems that consist of both operators' and users' decisions. The model finds stable pricing, in terms of generalized costs, and matches between user populations in a network to set of routes with line capacities. The proposed model gives a set of stable outcomes instead of single point pricing that allows operators to design ticket pricing, routes/schedules that impact access/egress, shared policies that impact wait/transfer costs, etc., based on a desired mechanism or policy. The set of stable outcomes is proven to be convex from which assignment-dependent unique user-optimal and operator-optimal outcomes can be obtained. Different user groups can benefit from using this model in a prescriptive manner or within a sequential design process. We look at several different examples to test our model: small examples of fixed transit routes and a case study using a small subset of taxi data in NYC. The case study illustrates how one can use the model to evaluate a policy that can require passengers to walk up to 1 block away to meet with a shared taxi without turning away passengers.

# 1. Introduction

Planning for mobility in a smart cities era requires an understanding beyond the route choices of travelers. With the increasing ubiquity of multiple forms of "Mobility as a Service" (MaaS) options to travelers (e.g. conventional fixed route transit, flexible transit, rideshare, carshare, microtransit, ridesourcing) provided by both public agencies and private operators known as "transportation network companies" (TNCs), travel forecast models need to focus on both the decisions of travelers and operators (Djavadian and Chow, 2017a). For example, a person's decision to take one mobility service over another may depend on the travel performance of that service, but the performance in turn depends on the operator's cost allocation decisions to best serve their users. Table 1 illustrates the broad range of cost allocation decisions exemplified by different types of mobility systems and how those decisions impacts costs for users and operators.

**Table 1**. Illustration of cost allocations

| Cost allocation | Cost transfer | Example systems |
| --- | --- | --- |
| Fare | User → Operator | Public transit, taxi, on-demand ridesharing, vehicle sharing |
| Wait time | Operator → User | Public transit, taxi, on-demand ridesharing |
| Access time | Operator → User | Public transit, vehicle sharing |
| Detour time | User → User | Public transit, on-demand ridesharing |
| Reservation time | Operator → User | Vehicle sharing, on-demand ridesharing |
| Capacity reliability | Operator → User | Public transit, vehicle sharing |
| Credit/discount for switching pickup/drop-off location | Operator → User | Public transit, on-demand ridesharing, vehicle sharing |
| Fare splitting | User → User | Public transit, on-demand ridesharing |

In this table, cost transfers refer to the direction of cost allocation: for example, a fare is a cost to a user that is transferred as a benefit to the operator. When planning for these systems, modelers need to forecast the outcomes of operators' cost allocation policies to forecast the route flows. The success and failure of various systems (e.g. Kutsuplus in Helsinki (Kelly, 2016), Car2Go in San Diego (Krok, 2016)) depend on forecasting the ridership, which is linked to the cost allocation decisions of those systems and the structure of other mobility options in their respective regions.

State-of-the-art techniques tend to keep these two decisions separate. For example, traffic or transit equilibrium models in general are designed without any operator response to obtain route flows that satisfy Wardrop user equilibrium principles. Using conventional transportation assignment methods, highly complex bilevel models with upper level Nash equilibrium are needed (e.g. Zhou et al., 2005). There are models to forecast certain equilibrium patterns like taxi-passenger matching (Yang et al., 2010) and ride-sourcing supply-demand equilibrium (Zha et al., 2016). The problem is that such methods do not allow city agencies to evaluate across multiple service types to compare alternatives and substitution effects between service designs.

System optimization models like vehicle sharing (Chow and Sayarshad, 2014) or ridesharing optimization (Masoud and Jayakrishnan, 2017; Wang et al., 2017) assume inelastic user demand. Those are generally normative models meant to be decision support for the operators, not as policy analysis tools for public agencies. An overview of this gap in the literature is given by Djavadian and Chow (2017a, b), who also propose simulation-based methods to evaluate such systems. However, a major drawback is that sensitivity analyses cannot be conducted, and the fundamental structure of the interactions is not clearly understood. For instance, the stability of a certain cost allocation strategy may be simulated, but the tool does not provide analytical thresholds to consider perturbations in a strategy or for transferability to other instances.

We propose a new transportation network assignment model framework based on stable matching for this purpose; it simultaneously considers user route behavior, service operator route selection decisions, and the resulting cost allocations and pricing mechanisms needed to reach a stable state. Having such a model allows operators to quantify the impacts of fleet operational algorithms in terms of user incentives, such as those requesting passengers to meet at pickup locations to reduce routing costs. Policymakers can also use such a model to analyze infrastructure policies that impact those operators, such as congestion pricing or allocated parking spaces for shared mobility services. Stable matching theory (Gale and Shapley, 1962) shows equilibrium between two disjoint sets (buyers and sellers). If transport services are the sellers and travelers are the buyers, finding their supply-demand equilibrium is a stable matching problem in a generalized sense. *However, the theory does not currently extend to matches made by travelers to links of a route.* This is needed to apply stable matching to route assignment models.

We generalize a prescriptive many-to-one assignment game to consider routes with multiple segments with line capacities for offline operating design analysis. The model formulation and solution method are also proposed to address stable matching of travelers to links of operator routes to help design systems where decision-makers have full or partial control of the assignment and/or cost allocation decisions. Variants of the model are proposed. We then demonstrate the applicability of the model to evaluate a public policy: if we explicitly consider user and operator incentives, what does the stable state for a shared taxi policy (Hu, 2017) look like in New York City? Our model can provide new insights that prior studies (e.g. Santi et al., 2014; Ma et al., 2017; Alonso-Mora et al., 2017) missed.

## 2. Review of assignment games

The stable matching problem has a long literature. Gale and Shapley (1962) first studied the problem through two applications: the "marriage problem" for one-to-one matching, and the "college admissions problem" for many-to-one matching. Shapley and Shubik (1971) formulated a linear program (LP) called the "assignment game" for matching problems that have transferable utilities. The "game" aspect refers to a cooperative game in which the splitting of the payoffs among the participants are made to ensure that they have sufficient incentives not to deviate. They showed that the region derived from the dual variables corresponding to the LP of the assignment game is "Core", where players do not have incentive to change their matched partner(s).

Consider two disjoint sets denoted by $P$ for buyers and $Q$ for sellers. A buyer $i \in P$ that matches with a seller $j \in Q$ earns a utility of $U_{ij}$. The item has a cost of production of $c_j$ for seller $j$. A successful match means the seller transfers the utility to the buyer with a price $p$. The difference between utility and cost of production can be interpreted as payoff $a_{ij} = \max(0, U_{ij} - c_j)$. Each buyer who satisfies her utility earns a profit from the difference between utility and the price of the item, $u_i = U_{ij} - p$. Each seller profits from the price sold minus the cost of production, $v_j = p - c_j$. The assignment game is essentially a game of outcome splitting. It is a cooperative game $(P, Q, a)$ wherein the players get payoffs by forming coalitions with each other. Each pair of $i$ and $j$ who make a coalition win a payoff with the value of $a_{ij}$. The goal in the assignment game is to match sellers and buyers in a way that the generated payoff from their coalitions is maximized. A review of the basic principles of the assignment game is provided in Roth and Sotomayor (1990). The assignment model formulation is shown here in Eq.(1)-(4).

$$\max \sum_{i \in P} \sum_{j \in Q} a_{ij} x_{ij} \qquad (1)$$

s.t.

$$\sum_{i \in P} x_{ij} \leq q_j \qquad \forall j \in Q \qquad (2)$$

$$\sum_{j \in Q} x_{ij} \leq w_i \qquad \forall i \in P \qquad (3)$$

$$x_{ij} \in \{0,1\} \qquad \forall j \in Q, \forall i \in P \qquad (4)$$

In the model, $x_{ij}$ is a binary variable which shows whether buyer $i$ and seller $j$ are matched or not. The parameters $q_j$ and $w_i$ are the quotas of each side. If $q_j$ and $w_i$ are equal to one, the assignment game becomes a one-to-one or simple assignment game. For any integer values of $q_j$ and $w_i$, the constraint set conforms to the Unimodularity Theorem, so the optimal solution of the LP relaxation is also integer.

In the simple assignment game, an outcome $((u, v); x)$ is feasible if $u_i$ and $v_j$ are non-negative and under a feasible assignment $x$ (i.e. $x$ satisfies Eq. (2) - (4)), $\sum_{i \in P} u_i + \sum_{j \in Q} v_j = \sum_{\substack{i \in P \\ j \in Q}} a_{ij} x_{ij}$. A feasible payoff is stable if $u_i + v_j = a_{ij}$ when $x_{ij} = 1$, and $u_i + v_j \geq a_{ij}$ when $x_{ij} = 0$. The core is the set of solutions of the dual corresponding to the assignment game. Note that this is not equivalent to Wardrop's user equilibrium and system optimal concepts which refer to competition between users when payoffs depend on their collective choices. Here we are looking at cooperation between buyers and sellers such that the outcome is consistent with behavior.

In the multiple partner assignment game (Sotomayor, 1992), the profit received by buyer $i$ (seller $j$) from matching to seller $j$ (buyer $i$) is defined as $u_{ij}$ ($v_{ij}$). The unmatched buyers and sellers are assumed to be matched to a dummy seller or buyer, respectively ($x_{ij_0}$ and $x_{i_0 j}$), with a payoff of zero. If $C(i, x)$ and $C(j, x)$ are defined as the set of matched players to $i$ and $j$, respectively, under optimal assignment of $x$, then $u_i$ is the minimum of $u_{ij}$ for each $j \in C(i, x)$ and $v_j$ is the minimum of $v_{ij}$ for each $i \in C(j, x)$. Stability of payoffs in the many-to-many assignment games implies that the feasible outcome $((u, v); x)$ is stable if $u_{ij} + v_{ij} = a_{ij}$ when $x_{ij} = 1$ and $u_i + v_j \geq a_{ij}$ when $x_{ij} = 0$, where $u_i$ and $v_j$ are non-negative for all $i$ and $j$. Furthermore, for the many-to-one game the stable matching outcomes correspond to the core of the assignment game such that there exists "buyer optimal" and a "seller optimal" outcomes.

The main challenge in a transportation network context is defining a many-to-one model structure where sellers are defined as routes with line capacities and multiple travelers are matched to segments of each route. We use the term "line" in reference to service lines (see Schöbel, 2012; Schiewe et al., 2019) on a route. For example, a route may serve 100 passengers in total but have a line capacity of 10 so that the number never exceeds that amount at any segment and direction, and three lines with the same line capacity may share an infrastructure link such that it cannot exceed 30 passengers on that link. This is the case with MaaS systems. Two different users can match to a route using two different node pairs but share the same line level "quotas" or capacities, all while maintaining a unique stable outcome space. These points are illustrated in more detail in Section 3.

While there are transportation applications using stable matching, cost allocation mechanisms, or cooperative game theory as shown in Table 2, none have considered both of the following: (1) assignment of travelers onto an operator route that can be composed of a sequence of nodes with line capacities, and (2) route-level cost allocation decisions of operators.

**Table 2**. Sample literature on transportation applications of stable matching, cost allocation mechanisms, and cooperative games.

| Reference | Type of network | Allocation decision | Mechanism |
|---|---|---|---|
| Bird (1976) | Minimum spanning tree | Core of Minimum spanning tree game | Set of stable allocations in core |
| Megiddo (1978) | Steiner tree | Demand nodes in Euclidian space | Core of Steiner tree |
| Kalai and Zemel (1982) | Maximum flow | Each player owns a single link | Core of maximum flow game |
| Derks and Tijs (1985) | Multicommodity flow | Each player owns a single link, different commodity | Core of multicommodity flow game |
| Curiel et al. (1989) | Maximum flow | Committee owns links and decide together | Core of maximum flow in centralized case |
| Granot and Granot (1992) | Maximum flow | N.A | N.A. |
| Matsubayashi et al. (2005) | Node pair | two types of cost, constant + coalition | Set of stable cost allocation in core |
| Potters et al. (2006) | Minimum cost flow | Players own link | Nucleolus of flow game |
| Agarwal and Ergun (2008) | Network | Players own capacities on links and also demands then cooperate | There is no cost allocation |
| Anshelevich et al. (2013) | Bipartite | Utility as a random value, cost of stability | N.A. |
| Dai and Chen (2015) | Nodes | Finding profit allocation inside core, efficiency measure | Profit allocation is done by egalitarian view |
| Stiglic et al. (2015) | Route | Finding meeting point with minimizing cost | Matching demand to routes through matching problem |
| Hezarkhani et al. (2016) | Route | Cooperative truck load delivery | Gain sharing is done in a procedure to be fare |
| Nourinejad and Roorda (2016) | Network | Decentralized and centralized models for dynamic ridesharing | Single-shot first-price Vickrey auction price |
| Aghajani and Kalantar (2017) | N.A. | Maximizing parking owner profit | Seller side optimal |
| Alonso-Mora et al. (2017) | Network | Algorithm for dynamic ridesharing matching and relocation of idle cars | No cost allocation |
| Hara and Hato (2017) | Node | Auction for car/bike sharing | Giving incentives to avoid imbalance problem |
| Masoud et al. (2017) | Node and routes (matching p2p ridesharing system) | Rider is charged for costs | Two types of cost: fixed cost (deviation from main route), variable cost (distance based) |
| Qian et al. (2017) | Network of NYC and taxi data | Incentives to taxi riders to share their cab | Best incentive that maximize efficiency of GRG |

| Rosenthal (2017) | Network | Cooperative game to cost allocating in rapid transit | Equally sharing |
| --- | --- | --- | --- |
| Wang et al. (2017) | Route | Dynamic ridesharing, matching individual cars to each other | Cost is divided equally between two matched pair |
| Lu and Quadrifoglio (2019) | Bipartite | Finding fair cost allocation for ridesharing services | They designed an algorithm to find Nucleolus |
| Papakonstantinou et al. (2019) | Network | Cooperative game between counties to mitigate sea level rise risk | Generated cost is divided fairly using Shapley Value |
| Peng et al. (2018) | Bipartite | Matching passengers and drivers with constraints | The pricing considering equity and incentive, is designed to ensure stability |

One area of the literature needs further elaboration. Coalition formation in networks is called a "network flow game". These studies deal with how multiple operators that own links in a network can form coalitions with each other to allow flows to occur. Examples include Bird's (1976) study for minimum spanning trees, Megiddo (1978) for Steiner trees, Kalai and Zemel (1982) for maximum flow problems, and Derks and Tijs (1985) for multicommodity flows. Modifications to the control scheme also exist. Curiel et al. (1989) allowed a group of operators to own a link with "committee control". Agarwal and Ergun (2008) allowed operators to own capacity on links instead of the whole links themselves. While network flow games also deal with cooperative games and coalition formation, the models are designed to analyze only interactions between different operators with each other. Primary applications include airline and freight industries.

On the contrary, the coalition formation in the proposed model is between travelers and the operators because the model is primarily designed to quantitatively analyze cost allocation policies between them, such as fare prices, reservation times, detours, and meeting points. This is fundamentally a different model framework than earlier "network flow games", even if both involve assignment games. In the case where travelers choose multiple operators to form a trip, a combination of both network flow game between multiple operators and user-to-operator stable matching may be required. This "many-to-many" extension will be studied in the future.

## 3. Proposed model

### 3.1. Definitions and model formulation

Unlike conventional transportation assignment models, the proposed model outputs not only traveler route flows and link performances, but also the set of stable cost allocations at the operator route level (fare prices, other generalized traveler cost transfers like additional walking or waiting time, etc.). It explicitly considers the incentive behavior of both travelers and operators. The model is used to evaluate service operating policies illustrated in Table 1 by determining stability of a policy and its cost allocation bounds for given demand patterns.

The most basic formulation setting is a static many-to-one (one operator route may match with multiple travelers, and one traveler is assigned to one route) assignment game. Consider a graph $G(N,A)$ in which there is a set $R$ of mobility operators' routes, where each route $r \in R$ is assumed to represent a separate "seller" and a set $S$ of user OD pairs looking for service. Each OD pair $s$ may include more than one traveler and be matched to multiple routes although each traveler individually can only be assigned to one route (integer solutions). A dummy user $k$ is created to match with routes that are not matched with any user. A route $r \in R$ is assumed to have only one sequence of links $a \in A_r$ between any pair of nodes served, where $A_r \subseteq A$ are disjoint sets. For example, for a 4-node network, one route may be 1-2-3-4, and another may be 3-2-1-4. A user that travels from node 1 to node 4 would have to visit 2, 3, and 4 if matched with the first route, and only node 4 if matched with the latter.

The following input parameters are needed. A match between an operator route $r$ and a user or set of homogenous users $s \in S$ imposes a generalized travel cost to the user(s), $t_{sr}$, that depends on the origin-destination (OD) of user $s$. Such a cost, for example, may be a function of multiple travel costs that include fares, wait time, access time, in-vehicle time, etc. This cost may be different for one user than another matched to the same route. For example, a user traveling from node 1 to node 4 on route 1-2-3-4 has a travel cost of visiting 2, 3, and then 4, whereas a traveler from node 2 to 3 on the same route would only incur a travel cost of visiting 3. The cost is fixed (there is no crowding effect on the line, just a hard capacity). Each user gains a utility $U_{sr}$ when matched to a route $r$ such that the net payoff is $a_{sr} = \max\{0, U_{sr} - t_{sr}\}$. When we set the utility for the next best travel option outside of the available options among $R$ to be zero, the $U_{sr}$ is defined so the payoff $a_{sr}$ represents the savings from that outside option. For a traveler with no other travel options (or if $R$ comprehensively includes all possible travel options), the next best travel option is simply to not make a trip in this system. Note that if user $s$ and route $r$ are not compatible pairs, we can model this with a very large travel cost ($t_{sr}$) so that it makes their payoff value ($a_{sr}$) equal to zero.

If we define $b_{sr}$ to be the minimum benefit acceptable for operator $r$ to match with user $s$, and $g_{sr}$ as the minimum acceptable benefit for user $s$ to match with route $r$, then the payoff value is $a_{sr} = \max\{0, U_{sr} - t_{sr} - g_{sr} - b_{sr}\}$. These minimum acceptable values just make the payoff values smaller and all the rest of characteristics of the proposed model remain same. Without a loss of generality, for the rest of this paper we assume the minimum acceptable profit for both users and operators is equal to zero.

Each route has an operating cost $C_r$ which requires the operator payoff allocation to exceed. For private operators, the fare payment portion of the allocation needs to exceed $C_r$ as some of the other user payoffs like travel time savings may not be transferable to offset operating cost. The cost of a route is divided between the users of that route ($c_{sr}$). Routes have line capacities defined as $w_r$. An indicator $\delta_{asr}$ is set to 1 if a match between user $s$ and route $r$ requires using link $a$, and 0 otherwise.

The output of the model is a set of matches $x_{sr}$ and the region of user profits $u_s$ and operator profits $v_r$ that are stable. For a given set of user and operator profits, there is a fare $p_s$ charged to the user by the matching route. The proposed model's formulation for single traveler OD pairs is shown in Eq. (5) - (9).

$$\max \sum_{s \in S} \sum_{r \in R} a_{sr} x_{sr} \tag{5}$$

s.t.

$$\sum_{r \in R} x_{sr} \leq 1 \qquad \forall s \in S \setminus \{k\} \tag{6}$$

$$\sum_{s \in S \setminus \{k\}} \delta_{asr} x_{sr} \leq w_r \qquad \forall a \in A_r, r \in R \qquad (7)$$

$$\sum_{s \in S \setminus \{k\}} x_{sr} \leq M(1 - x_{kr}) \qquad \forall r \in R \qquad (8)$$

$$x_{sr} \in \{0,1\} \qquad \forall s \in S, r \in R \qquad (9)$$

Eq. (5) is the standard assignment game objective to maximize payoffs. Eq. (6) and (7) correspond to the matching quotas for a many-to-one assignment game. The line capacity for the route operators is met on their lines. Eq. (8) is a new constraint that ensures the routes are only matched if the sum of all the matched payoff allocations exceeds the operating cost. Eq. (9) are the integral constraints. By setting $a_{kr} = C_r$, $\forall r \in R$, a route would only be matched if the total payoff (in generalized user utility converted to operating dollars) exceeds the operating cost because of Eq. (8).

For the cases that have a set $H$ of user (OD) demand greater than one, we can easily set the right side of Eq. (6) equal to $q_s$ as the demand value and consider $x_{sr}$ as an integer ($x_{sr} \in \mathbb{Z}_+$). While this may seem like a many-to-many system since one OD pair may be matched to multiple routes, behaviorally it is still a many-to-one system because the matching condition is for each traveler to be matched to one route. For the stability of such cases we can simply separate the user bundle $s$ to $|H|$ single user (OD) and treat each of them as a single independent agent (integer solution).

Figure 1 illustrates the model setting; based only on travel costs (as typically done in traffic assignment models) "Feasible Output 2" seems to be the preferred choice, but based on the choices of other passengers, the operating cost $C_r$, and the savings from other travel options, "Feasible Output 1" might be the chosen one.

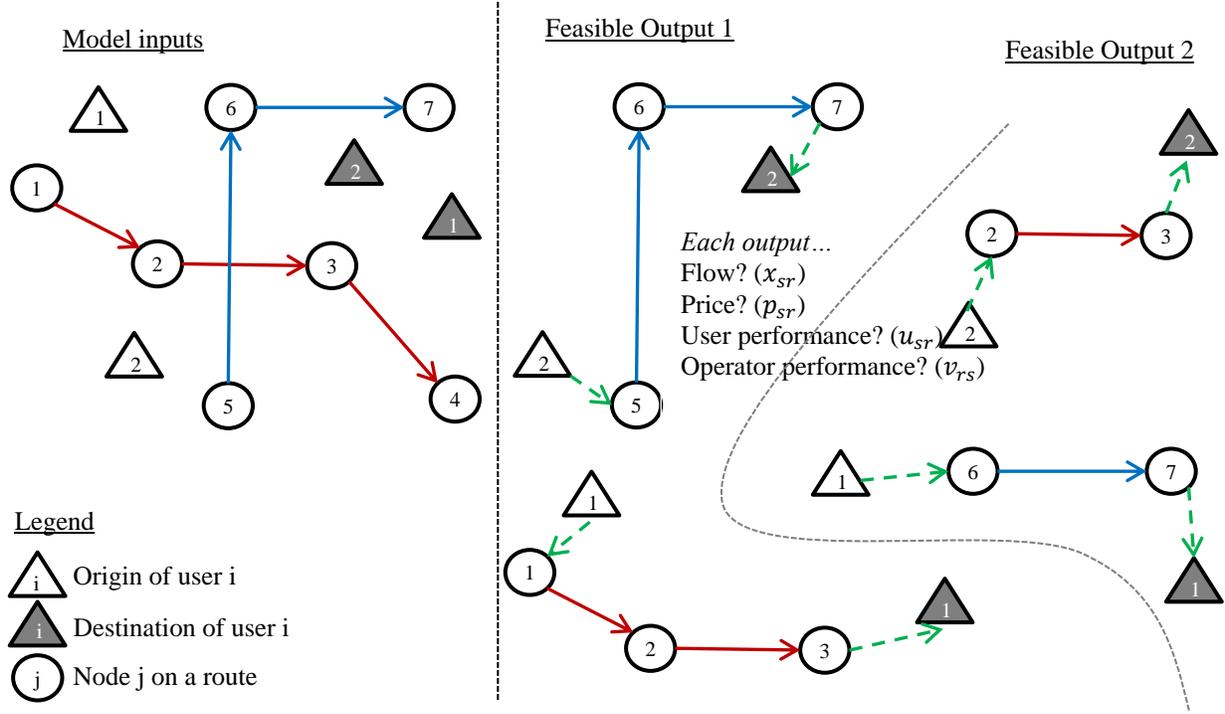

**Figure 1.** Illustration of explaining different assignment configurations using the proposed model.

The payoff value used in the objective function in Eq. (5) is defined as $a_{sr} = \max\{0, U_{sr} - t_{sr}\}$, where $U_{sr}$ is the utility attained by the user for matching with a route, converted to units of dollars, and $t_{sr}$ is a generalized travel disutility to the user that can include monetary values of wait time, access time, or any other costs not transferred to the operator. The travel disutility may take any functional form as best fit to the data. The classic payoff value defined in Shapley and Shubick (1971) is $a_{sr} = \max\{0, U_{sr} - t_{sr} - c_{sr}\}$, where $c_{sr}$ is the cost of operating the route attributed to the user. The issue is that $c_{sr}$ is an endogenous variable that is a function of $x_{sr}$ and a cost sharing mechanism; for example, two users sharing a ride can divide the cost of providing that trip as opposed to one user paying for the total cost alone. We claim that our model is equivalent to the case in Shapley and Shubik (1971) under Assumption 1, which only requires that the cost sharing mechanism be equal to any sum between users. We show this in Proposition 1.

**Assumption 1**. *Operating cost $C_r$ can be divided among different $c_{sr}$ for each user $s$ that is matched to $r$, i.e. there exists a set of $c_{sr}$ such that $C_r = \sum_{s \in S \setminus \{k\}} c_{sr} x_{sr}$.*

**Proposition 1.** The objective function of model (5) – (9), ($a_{sr} = \max\{0, U_{sr} - t_{sr}\}$), is equivalent to the objective function of the classic assignment model that defined in Shapley and Shubik (1971) with explicitly known production costs ($a_{sr} = \max\{0, U_{sr} - t_{sr} - c_{sr}\}$)..

**Proof.**
Objective function of model (5) - (9) can be written as Eq.(10).

$$\max \sum_{s \in S} \sum_{r \in R} a_{sr} x_{sr} = \max \left( \sum_{s \in S \setminus \{k\}} \sum_{r \in R} (U_{sr} - t_{sr}) x_{sr} + \sum_{r \in R} C_r x_{kr} \right) \quad (10)$$

The right hand side of Eq. (10) consists of two parts. The second part is the total route cost of unused routes. Let's define $\bar{R} \subset R$ as the set of used routes. Then the second part of Eq. (10) can be written as Eq. (11).

$$\sum_{r \in R} C_r x_{kr} = \sum_{r \in R} C_r - \sum_{r \in \bar{R}} C_r \quad (11)$$

Now the objective function of the main model (Eq. (5)) can be rewritten again as Eq. (12).

$$\max \sum_{s \in S} \sum_{r \in R} a_{sr} x_{sr}$$
$$= \max \left( \sum_{s \in S \setminus \{k\}} \sum_{r \in R} (U_{sr} - t_{sr}) x_{sr} + \sum_{r \in R} C_r - \sum_{r \in \bar{R}} C_r \right) \quad (12)$$

The term $\sum_{r \in R} C_r$ is constant and can be eliminated from objective function (12). Based on what is defined, $\sum_{r \in \bar{R}} C_r$ can be written as $\sum_{r \in R} \sum_{s \in S \setminus \{k\}} c_{sr} x_{sr}$. Eq. (12) becomes Eq.(13).

$$\max \left( \sum_{s \in S \setminus \{k\}} \sum_{r \in R} (U_{sr} - t_{sr}) x_{sr} - \sum_{r \in R} \sum_{s \in S \setminus \{k\}} c_{sr} x_{sr} \right)$$
$$= \max \left( \sum_{s \in S \setminus \{k\}} \sum_{r \in R} (U_{sr} - t_{sr} - c_{sr}) x_{sr} \right) \quad (13)$$

This net payoff is equivalent to Shapley and Shubik (1971). ∎

The utility parameter $U_{sr}$ needs to be accurately calibrated in advance. For user groups with $q_s > 1$, this parameter should be representative of a population. The value depends on the presence of other transport options outside of the set of $R$ being considered, and on the types of trip purposes at the destination. For example, the utility of a non-compulsory trip may be compared primarily to not making a trip at all, whereas a compulsory work trip may have a utility based on comparing against another travel mode even if only walking is available. Parameter estimation for mathematical programming models can be done using inverse optimization (Ahuja and Orlin, 2001; Xu et al., 2018).

### 3.2. Stable matching properties
The optimal solution to the assignment game determines the assignment of users and operators to each other as well as the cost allocation space. Let's consider a cost transfer (e.g. fare price, among others shown in Table 1) from user $s$ to operator route $r$, $p_{sr}$, and $v_{rs}$ be the profit that operator $r$ earns from serving user $s$. By definition, $p_{sr} = c_{sr} + v_{rs}$. Let $u_s$ be the value gained by user $s$ from matching.

Cost allocation between the users and operators should lead to a stable outcome. Before starting to define the stability criteria, let us define the notation used. We define $D(r, x)$ as the set of users that are unmatched to $r$ under $x$ assignment, $C(r, x)$ as the set of users that are matched to $r$ (except the unmatched quota we assumed that are matched to null user) ($D(r, x) \cup C(r, x) = S \setminus \{k\}$). The same definitions hold for $D(s, x)$ and $C(s, x)$. As defined earlier $\bar{R}$ is the set of routes that are matched to at least one user. $v_r = \sum_{s \in C(r,x)} v_{rs}$ is the total benefit that route $r$ gains from its matches in assignment $x$.

**Definition 1.** (Sotomayor, 1992) *Outcomes $u_s$ and $v_r$ are feasible and denoted by $((u, v); x)$ if:*
  (i)   $\sum_{s \in C(r,x)} u_s + v_r = \sum_{s \in C(r,x)} a_{sr} - C_r$ and $u_s \geq 0, v_r \geq 0$ $\forall r \in \bar{R}$
  (ii)  $v_r = 0$ $\forall r \in R \setminus \bar{R}$
  (iii) $u_s = 0$ $\forall s$ such that $C(s, x) = \emptyset$

**Definition 2. (Core)** *A feasible outcome $((u, v), x)$ of the assignment game in Eq. (5) - (9) is stable if it satisfies Eq. (14).*

$$\sum_{s \in G_r} u_s + v_r \geq \sum_{s \in G_r} a_{sr} - C_r \qquad \forall G_r \text{ and } r \in R \qquad (14)$$

*where $G_r$ is set of user groups that can be feasibly matched to route $r$ (i.e. satisfies constraint of Eq. (7)).*

This stability condition ensures that no other coalition of users can make a better payoff than the current assignment solution and outcome allocation. Based on the definitions, the amount of generalized cost that users transfer to operators, $p_{sr}$, depends on the values allocated to operators, $v_{rs}$, and the $a_{sr}$ which is obtained from the utilities.

To clarify the definitions of stability and feasibility of outcomes, we show them in a small example shown in Figure 2.

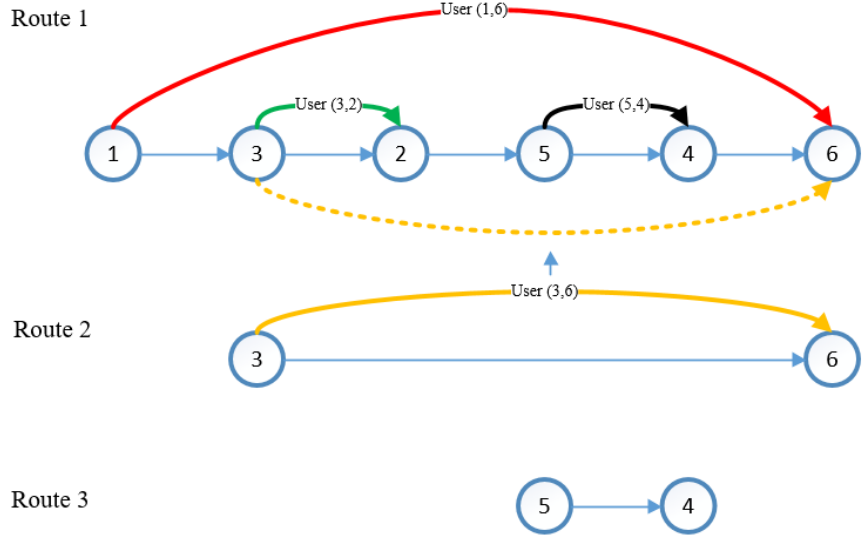

**Figure 2.** Users and routes interaction in stability.

As shown in Figure 2, suppose there is a set of three routes $\{r_1 = (1 \to 3 \to 2 \to 5 \to 4 \to 6), r_2 = (3 \to 6), r_3 = (5 \to 4)\}$ where each has a quota of 2 on each of its links, and a set of four users $S = \{(1,6), (3,2), (5,4), (3,6)\}$. Under assignment $x$, $s_1, s_2, s_3$ and $s_4$ are respectively matched to routes $r_1, r_1, r_1$ and $r_2$. Then we have: $C(r_1, x) = \{s_1, s_2, s_3\}$, $C(r_2, x) = \{s_4\}$, $C(r_3, x) = \emptyset$ and $D(r_1, x) = \{s_4\}$, $D(r_2, x) = \{s_1, s_2, s_3\}$, $D(r_3, x) = \{s_1, s_2, s_3, s_4\}$. The equations for stability and feasibility conditions for this example are shown below.

<div align="center">Conditions for Feasibility</div>

$$(u_1 + u_2 + u_3) + v_1 = (a_{11} + a_{21} + a_{31}) - C_1$$
$$u_4 + v_2 = a_{42} - C_2$$
$$v_3 = 0$$
$$u_1, u_2, u_3, u_4, v_1, v_2, v_3 \geq 0$$

<div align="center">Conditions for Stability</div>

$$(u_1) + v_1 \geq (a_{11}) - C_1 \qquad (u_2 + u_3) + v_1 \geq (a_{21} + a_{31}) - C_1$$
$$(u_2) + v_1 \geq (a_{21}) - C_1 \qquad (u_1 + u_2 + u_3) + v_1 \geq (a_{11} + a_{21} + a_{31}) - C_1$$
$$(u_3) + v_1 \geq (a_{31}) - C_1 \qquad (u_1 + u_4) + v_1 \geq (a_{11} + a_{41}) - C_1$$
$$(u_1 + u_2) + v_1 \geq (a_{11} + a_{21}) - C_1 \qquad (u_4) + v_1 \geq (a_{41}) - C_1$$
$$(u_1 + u_3) + v_1 \geq (a_{11} + a_{31}) - C_1 \qquad (u_3) + v_3 \geq (a_{33}) - C_3$$

**Proposition 2.** *The set of stable outcomes for the assignment game in Eq. (5) – (9) consists of a convex space.*

**Proof.**
The proof follows directly from the fact that the set of stable outcomes is defined by a set of linear constraints (Eq. (14) and feasibility definition) so the space of stable outcomes is convex. ∎

We can now establish the necessary and sufficient conditions to relate an optimal assignment to a stable outcome with Proposition 3. Furthermore, it is known assignments and even the dual variables corresponding to the line capacities can be non-unique (e.g. Larsson and Patriksson, 1999); we show that non-unique assignments nonetheless share the same stable outcome space with Proposition 4.

**Proposition 3.** *Assignment $x$ corresponding to stable outcome $\big((u,v);x\big)$ satisfying Eq. (14) is an optimal assignment solution to the assignment problem in Eq. (5) – (9).*

**Proof.**
Stability of outcomes in assignment $x$ with respect to any other assignment $x'$ implies that

$$\sum_{s \in C(r,x')} u_s + v_r \geq \sum_{s \in C(r,x')} a_{sr} - C_r = \sum_{s \in S \setminus \{k\}} a_{sr} x'_{sr} - C_r \qquad \forall r \in R \qquad (15)$$

Summation of Eq. (15) over all the $r \in \bar{R}$ in $x'$ results in Eq. (16).

$$\begin{aligned}\sum_{s \in S \setminus \{k\}} u_s + \sum_{r \in \bar{R}} v_r &\geq \sum_{s \in S \setminus \{k\}} \sum_{r \in \bar{R}} a_{sr} x'_{sr} - \sum_{r \in \bar{R}} C_r \\ &= \sum_{s \in S \setminus \{k\}} \sum_{r \in R} a_{sr} x'_{sr} - \sum_{r \in R} (1 - x'_{kr}) C_r \end{aligned} \qquad (16)$$

Using the feasibility definition, the left side of Eq. (16) can be rewritten as Eq. (17).

$$\begin{aligned}\sum_{s \in S \setminus \{k\}} u_s + \sum_{r \in \bar{R}} v_r &= \sum_{s \in S \setminus \{k\}} u_s + \sum_{r \in R} v_r \\ &= \sum_{s \in S \setminus \{k\}} \sum_{r \in R} a_{sr} x_{sr} - \sum_{r \in R} (1 - x_{kr}) C_r \end{aligned} \qquad (17)$$

By inserting Eq. (17) into Eq. (16) we have Eq. (18).

$$\sum_{s \in S \setminus \{k\}} \sum_{r \in R} a_{sr} x_{sr} - \sum_{r \in R} (1 - x_{kr}) C_r \geq \sum_{s \in S \setminus \{k\}} \sum_{r \in R} a_{sr} x'_{sr} - \sum_{r \in R} (1 - x'_{kr}) C_r \qquad (18)$$

Eq. (18) can be re-written in the form of Eq. (19).

$$\sum_{s \in S} \sum_{r \in R} a_{sr} x_{sr} \geq \sum_{s \in S} \sum_{r \in R} a_{sr} x'_{sr} \qquad (19)$$

This means the objective function of assignment $x$ is better than any other assignment $x'$ which shows the optimality of assignment $x$. ∎

**Proposition 4.** *Any optimal assignment solution $x$ to the assignment problem in Eq. (5) – (9) shares the same stable outcome area with every other optimal assignment solution $x'$.*

**Proof.**
Let's say we have two optimal assignment solutions $x$ and $x'$. If we show that the stable outcome $((u,v);x)$ is generally feasible for assignment $x'$ then we can say that $((u,v);x)$ is also the stable space for assignment $x'$.

From the stability condition we have Eq. (15) and summing up over all the routes $r$ we get the Eq. (16). Since the $x'$ is also an optimal solution, the Eq. (16) will be in the form of equality. This means that all the Eq. (15) also have the form of equality necessary for the feasibility of $((u,v);x)$ for assignment $x'$. ∎

Lastly, we show with Proposition 5 that the stability conditions extend to a generalized assignment problem with multiple travelers per OD pair, i.e. when $q_s > 1$. The proposition clarifies the distinction, from individual behavior in which users select different routes, to the Wardrop's equilibrium that arises when identical travelers end up choosing different routes. In essence, the routes are chosen only if they are equivalent in value when considering the dual values of the link capacities. This proposition also maintains that the problem is strictly a many-to-one assignment game cast within a capacitated network context.

**Proposition 5.** *Identical users (like user demand sharing the same OD when $q_s > 1$ for the user bundle version of assignment game in Eq. (5) – (9)) gain equal benefit from their match to different routes.*

**Proof.**
In Proposition 4 we saw that all the stable outcomes are sharing the same stable outcome space. Let's say the user bundle $s$ includes two agents $e$ and $e'$. In an optimal assignment solution $x$, $e$ is matched to route $r$ and $e'$ to $r'$ and in the other optimal assignment solution $x'$, $e$ is matched to route $r'$ and $e'$ to $r$. Note that two agents $e$ and $e'$ are completely identical. Stability conditions for assignment $x$ implies Eq. (20) – (23).

$$u_e + \sum_{s \in C(r,x)\setminus\{e\}} u_s + v_r = a_{er} + \sum_{s \in C(r,x)\setminus\{e\}} a_{sr} - C_r \qquad (20)$$

$$u_e + \sum_{s \in C(r',x)\setminus\{e'\}} u_s + v_{r'} \geq a_{er'} + \sum_{s \in C(r',x)\setminus\{e'\}} a_{sr'} - C_{r'} \qquad (21)$$

$$u_{e'} + \sum_{s \in C(r',x)\setminus\{e'\}} u_s + v_{r'} = a_{e'r'} + \sum_{s \in C(r',x)\setminus\{e'\}} a_{sr'} - C_{r'} \qquad (22)$$

$$u_{e'} + \sum_{s \in C(r,x)\setminus\{e\}} u_s + v_r \geq a_{e'r} + \sum_{s \in C(r,x)\setminus\{e\}} a_{sr} - C_r \qquad (23)$$

Stability condition for assignment $x'$ implies Eq. (24) – (27).

$$u_e + \sum_{s \in C(r',x')\setminus\{e\}} u_s + v_{r'} = a_{er'} + \sum_{s \in C(r',x')\setminus\{e\}} a_{sr'} - C_{r'} \qquad (24)$$

$$u_e + \sum_{s \in C(r,x') \setminus \{e'\}} u_s + v_r \geq a_{er} + \sum_{s \in C(r,x') \setminus \{e'\}} a_{sr} - C_r \tag{25}$$

$$u_{e'} + \sum_{s \in C(r,x') \setminus \{e'\}} u_s + v_r = a_{e'r} + \sum_{s \in C(r,x') \setminus \{e'\}} a_{sr} - C_r \tag{26}$$

$$u_{e'} + \sum_{s \in C(r',x') \setminus \{e\}} u_s + v_{r'} \geq a_{e'r'} + \sum_{s \in C(r',x') \setminus \{e\}} a_{sr'} - C_{r'} \tag{27}$$

Since agents $e$ and $e'$ have the same characteristics, $a_{er} = a_{e'r}$ and $a_{er'} = a_{e'r'}$. From this and Eq. (20) - (27) we can say that $u_e = u'_e$. ∎

### 3.3. Solution method

The proposed model is conveniently cast as an integer programming problem with quota constraints, but with route flows for different OD pairs and line capacities. The model is a type of capacitated multicommodity flow problem, where the latter is known to be NP-hard (Even et al., 1975) with unsplittable or discrete flows. Any conventional IP solution algorithm can be applied to solve this model and LP relaxation can be applied when allowing for continuous flows. There are several approaches which can be divided into three main categories of methods: cutting planes (e.g. reducing feasible area by adding extra constraints), heuristic methods (e.g. search methods, subgradient optimization, see Held et al., 1974), and implicit enumeration techniques (e.g. Branch and Bound, see Lawler and Wood, 1966).

Having found a set of assignments, a stable cost sharing problem is expressed as Eq. (28) to (33). The decision variables are $u_s$ and $v_r$. Constraint (29) is related to definition of stability. The objective function $\max Z$ represents a desired cost allocation mechanism, such as maximizing revenue, social welfare, fairness, etc.

$$\max Z \tag{28}$$

s.t.

$$\sum_{s \in G_r} u_s + v_r \geq \sum_{s \in G_r} a_{sr} - C_r \qquad \forall G_r \text{ and } r \in R \tag{29}$$

$$\sum_{s \in C(r,x)} u_s + v_r = \sum_{s \in C(r,x)} a_{sr} - C_r \qquad \forall r \in \bar{R} \tag{30}$$

$$v_r = 0 \qquad \forall r \in R \setminus \bar{R} \tag{31}$$

$$u_s = 0 \qquad \forall s \text{ s.t. } C(s,x) = \emptyset \tag{32}$$

$$u_s \geq 0, v_r \geq 0 \qquad \forall r \in \bar{R} \tag{33}$$

When $Z$ is set as either $\sum_{s \in S} u_s$ or $\sum_{r \in R} v_r$, the solutions obtain the user-optimal and operator-optimal outcomes, respectively. These two objective functions form opposite vertices of the space of the set of stable outcomes. Another objective function may obtain an outcome allocation that lies somewhere between these two bounds. Different cost allocation mechanisms may be used, although some may not result in allocations within this stable outcome space. By setting a desired mechanism we can use Eq. (28) – (33) to set prices for the assignment. When $Z$ is linear, the resulting model is a simple LP for obtaining the cost allocations under the stable matching.

The stable cost allocation of this study is based on the definition of the core. A core is a strong stability condition that leads to two issues. First, as illustrated with the example in Figure 2, all the possible coalitions need to be checked so that they cannot generate more payoff than what they get in the core (Eq. (29)). Generation of all possible coalitions can be computationally expensive. On the other hand, the core may not always be non-empty. This can be advantageous in the design of a system because one that results in an empty core suggests there is no advantage for the operators to serve the users. In the next section, we show a numerical example where the core has different conditions with different parameters.

The computational complexities of the assignment game and core allocation depend on the number of routes enumerated. Implicit enumeration techniques are the most obvious way to improve computational efficiency for large scale application. For example, Barnhart et al. (1994) proposed using column generation for implicit route enumeration to tackle large scale problems which we will investigate further in future research.

### 3.4. Illustration with fixed route transit service
### 3.4.1. Network parameters

To illustrate how this model works, we present a simple 4-node transit route network example shown in Figure 3. The number over each link in the figure represents the travel time between the nodes. In this network, we allow for all possible routes to be enumerated, of which there are 52 candidate routes ($R = \{r_1, \ldots, r_{52}\}$). We consider a cost structure based on the number of links in the route (a route with more links has higher operating cost), which is set to $C_r = 4.5 + 0.5 \times |A_r|$.

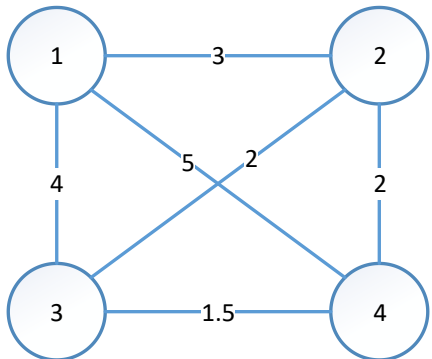

**Figure 3.** 4-node network example.

Two different problem settings are considered. The first illustrates individuals with binary decisions where no capacity is assumed ($w_r \to \infty$). In the second case, for each "user" there is demand of 5 individuals for each OD pair. The line capacity is set to 2 in this latter case. These examples can be interpreted as transit routes that serve a population of travelers. In the computational experiments we set prices based on user-optimal ($Z = \sum_{s \in S \setminus \{k\}} u_s$) and operator-optimal ($Z = \sum_{r \in R} v_r$) mechanisms. Commercial optimization packages use one or combine some of these methods to solve the integer programming problems. For convenience we used Matlab's *intlinprog* solver for the numerical examples in this study, as it uses a variant of the branch and bound algorithm to obtain a solution.

### 3.4.2. Binary, non-capacitated case

A set of demand OD pairs between some of the OD pairs are randomly chosen ($S = \{(1,2),(1,3),(2,3),(3,2),(4,1),(4,2)\}$) for this experiment. The utility ($U_{rs}$) for conducting each trip for all users is set constant ($U_{sr} = U = 20 \quad \forall s \in S\setminus\{k\}, r \in R$).

After solving the assignment model of Eq. (5) - (9), the optimal assignments is shown in Table 3a and Figure 4. The objective value for the optimal assignment solution is 88.5. Solving the LP in Eq. (28) to (33) with $Z$ equal to $\sum_{s \in S} u_s$ and $\sum_{r \in R} v_r$ results in the payoffs for user-optimal and operator-optimal cases as the ticket price relating to these two optimal payoffs are shown in Table 3b and 3c. The $Z^*$ for user- and operator-optimal objective values are 88.5 and 0.5 respectively.

**Table 3a.** Result of assignment game for four node network

| | $r$ | | | Users (O,D) | | | | |
|---|---|---|---|---|---|---|---|---|
| number of route | Links of route | Cost of route | (1,2) | (1,3) | (2,3) | (3,2) | (4,1) | (4,2) |
| 29 | 1 − 3 − 4 − 2 | 6 | | * | | * | | * |
| 51 | 4 − 1 − 2 − 3 | 6 | * | | * | | * | |

**Table 3b.** Ticket prices in a user-optimal allocation mechanism

| Route | Links of route | Cost of route | User ticket prices (O,D) | | | | | | Operator revenue |
|---|---|---|---|---|---|---|---|---|---|
| | | | (1,2) | (1,3) | (2,3) | (3,2) | (4,1) | (4,2) | |
| 29 | 1 − 3 − 4 − 2 | 6 | | 0.98 | | 2.42 | | 2.60 | 6 |
| 51 | 4 − 1 − 2 − 3 | 6 | 2.24 | | 2.79 | | 0.97 | | 6 |

**Table 3c.** Ticket prices in an operator-optimal allocation mechanism

| Route | Links of route | Cost of route | User ticket prices (O,D) | | | | | | Operator revenue |
|---|---|---|---|---|---|---|---|---|---|
| | | | (1,2) | (1,3) | (2,3) | (3,2) | (4,1) | (4,2) | |
| 29 | 1 − 3 − 4 − 2 | 6 | | 1 | | 2.39 | | 2.62 | 6.01 |
| 51 | 4 − 1 − 2 − 3 | 6 | 2.65 | | 2.85 | | 0.98 | | 6.48 |

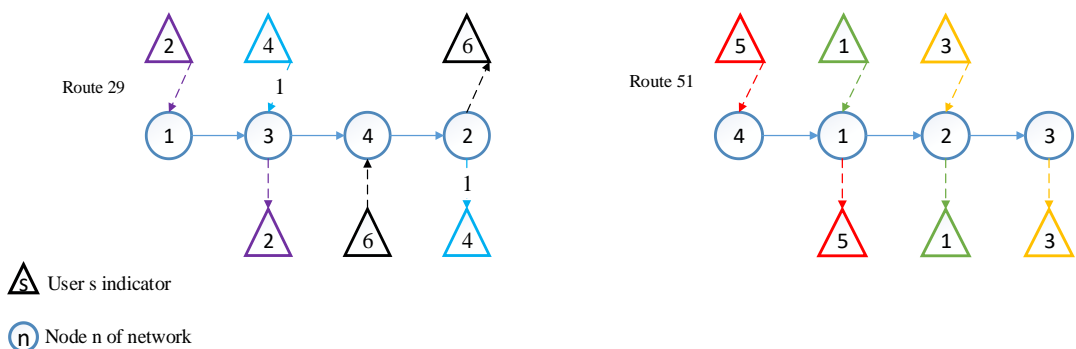

**Figure 4.** 4-node example user route matching in different optimal solutions.

Based on feasibility of outcomes, matched partners have the following payoffs: $u_2 + u_4 + u_6 + v_{29} = 44.5; u_1 + u_3 + u_5 + v_{51} = 44$. Both $b_{rs}$ and $g_{sr}$ are assumed to be zero.

The result for payoff splitting between the players is shown in Table 3b and 3c. In the user-optimal allocation, the ticket prices are as small as possible and the operators gain no profit in this matching. On the contrary, in the operator-optimal allocation, the prices are set as high as possible.

Since the set of stable outcomes is convex, every other ticket price between these two ticket prices are also stable ticket prices.

### 3.4.3. Integer, capacitated case

In this scenario, users from the prior setting are modified into user bundles with $q_s = 5 \; \forall s \in S\setminus\{k\}$. A line capacity is added to each route as well, where $w_r = 2, \; \forall r \in R$.

The result of the assignment game is shown in Table 4 and Figure 5. In Figure 5 the number of passengers is written on the edge from user to network node. Due to capacity, some users switch to other operator routes. For this problem with these parameters, the set of stable outcomes is empty (Core is empty). It means that constraints of Eq. (29) to (33) do not contain any feasible area.

Now consider a user group size of $q_s = 2 \; \forall s \in S\setminus\{k\}$ instead. The assignment solution is shown in Table 5a. The optimal objective value is 187.5. The operator-optimal and user-optimal allocations from Eq. (28) to (33) for this case are shown in Table 5b, illustrating the stable outcome under operator- and user-optimal mechanisms. The value of $Z^*$ for user- and operator-optimal objectives are 93.75 and 0 respectively. In this scenario, the outcomes for both the operator- and user-optimal mechanisms are identical which means the core is a single point.

**Table 4.** Result of assignment for 4-node network with demand ($q_s = 5 \; \forall s \in S\setminus\{k\}$) and route link capacity

| | $r$ | | User group (O,D) | | | | | |
|---|---|---|---|---|---|---|---|---|
| route number | Links of route | Cost of route | (1,2) | (1,3) | (2,3) | (3,2) | (4,1) | (4,2) |
| 6 | 4 − 2 | 5 | | | | | | 2 |
| 7 | 1 − 3 − 2 | 5.5 | | 2 | | 2 | | |
| 9 | 1 − 2 − 3 | 5.5 | 2 | | 2 | | | |
| 25 | 4 − 1 − 2 | 5.5 | 2 | | | | 2 | |
| 28 | 4 − 2 − 3 | 5.5 | | | 2 | | | 2 |
| 29 | 1 − 3 − 4 − 2 | 6 | | 1 | | 1 | | 1 |
| 49 | 4 − 1 − 3 − 2 | 6 | | 2 | | 2 | 2 | |
| 51 | 4 − 1 − 2 − 3 | 6 | 1 | | 1 | | 1 | |
| | Total | | 5 | 5 | 5 | 5 | 5 | 5 |

**Table 5a.** Result of assignment for 4-node network with demand ($q_s = 2 \; \forall s \in S\setminus\{k\}$) and route link capacity

| | $r$ | | User group (O,D) | | | | | |
|---|---|---|---|---|---|---|---|---|
| route number | Links of route | Cost of route | (1,2) | (1,3) | (2,3) | (3,2) | (4,1) | (4,2) |
| 1 | 1 − 2 | 5 | 2 | | | | | |
| 28 | 4 − 2 − 3 | 5.5 | | | 2 | | | 2 |
| 49 | 4 − 1 − 3 − 2 | 6 | | 2 | | 2 | 2 | |
| | Total | | 2 | 2 | 2 | 2 | 2 | 2 |

**Table 5b.** Ticket prices in operator-optimal and User-optimal allocation mechanisms

| Cost allocation mechanism | Route | Links of route | Cost of route | User group (O,D) | | | | | | Operator revenue |
|---|---|---|---|---|---|---|---|---|---|---|
| | | | | (1,2) | (1,3) | (2,3) | (3,2) | (4,1) | (4,2) | |
| | 1 | 1 − 2 | 5 | 2.5 | | | | | | 5 |

| | | | | | | | | | |
|---|---|---|---|---|---|---|---|---|---|
| User optimal | 28 | $4-2-3$ | 5.5 | | | 0.25 | | | 2.5 | 5.5 |
| | 49 | $4-1-3-2$ | 6 | | 0.46 | | 2.29 | 0.25 | | 6 |
| Operator optimal | 1 | $1-2$ | 5 | 2.5 | | | | | | 5 |
| | 28 | $4-2-3$ | 5.5 | | | 0.25 | | | 2.5 | 5.5 |
| | 49 | $4-1-3-2$ | 6 | | 0.46 | | 2.29 | 0.25 | | 6 |

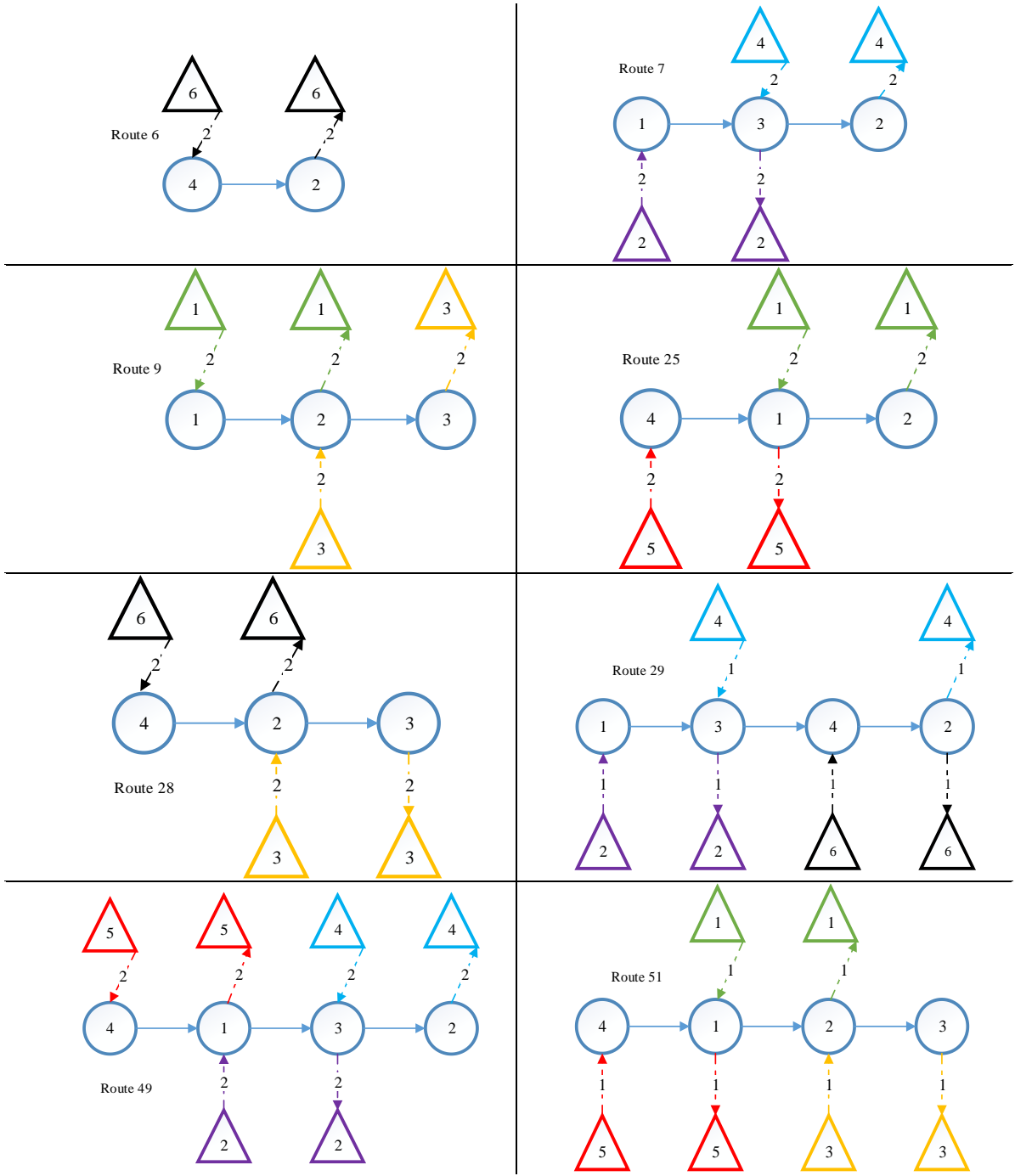

User s indicator

Node n of network

**Figure 5.** Matching of users and network nodes in 4-node network example with capacity.

This test demonstrates how the stable matching assignment game can be applied to OD-level population groups in a network of routes that exhibit line capacities. The results show that a stable matching can be found along with a corresponding unique stable outcome space for determining stable cost allocations. Moreover, it shows how the core is dependent on different parameters of an example.

## 4. Case study

### *4.1. Experimental design*
Having demonstrated how the model works, we now apply it to a case study calibrated to real data to illustrate the model in a more realistic setting as a proof-of-concept. We choose a taxi case study instead of a public transit one for several reasons:
1) Taxi ridership data is readily available through the NYC Open Data portal.
2) Taxi trips represent individual trips, and recent news about allowing shared rides (Hu, 2017) suggests the question of demand for shared rides remains an open question that is of interest to policymakers.
3) Tools to address this question have looked at either simulations that assume inelastic demand (Santi et al., 2014) or considered demand only for very specific trip purposes like airport access (Ma et al., 2017). Equilibrium assignment that considers both taxi operator and user incentives has not been conducted.

We choose to study this problem as follows. A random sample of taxi riders is selected based on pickups and drop-offs during a particular time period in lower Manhattan. Two scenarios are defined; one in which only single rider taxi routes are used, and one in which ridesharing is an option. The single rider taxi data are used to calibrate the payoff values. We assume utilities are equal to the cost of the trips made in that base scenario. The assignment game model is expected to inform on:
- Percent of users who choose to share rides considering incentives
- Price allocations under two schemes: user-optimal and taxi-optimal allocation
- Decision support for designing a cost allocation policy to request travelers meet the vehicles at a common distance

Taxi operations are dynamic whereas our proposed model is a static model. To deal with this discrepancy, we carefully calibrate our base case such that representing the observed data with a static, multiperiod assignment would be stable. For more rigorous treatment of dynamic assignment, non-myopic assignment considerations with potential spillovers of requests (e.g. Sayarshad and Chow, 2015) and online cost allocations (Furuhata et al., 2015) would be needed, which are beyond the scope of this study. While this simplification and the small sample size prevent us from drawing rigorous empirical conclusions about the shared taxi policy across the whole NYC, we can at least use the experiment to illustrate how the model can be used.

*4.2. Data*

NYC Taxi and Limousine Commission releases data every month on taxi trips in NYC. The data provides valuable information about each trip, such as pick up and drop off times and locations, trip distance, number of passengers, payment type and detailed fare data (fare + tax + tip). For this example, taxi data from Wednesday October 5th 2016 from 8AM to 8:30AM was used. There are 19,972 taxi trips made in NYC during this time. We consider the lower Manhattan region below 23rd Street as our study area, as shown in Figure 6. We create 21 nodes to represent zone centroids. Distance and travel time matrices between the nodes are extracted from Google API.

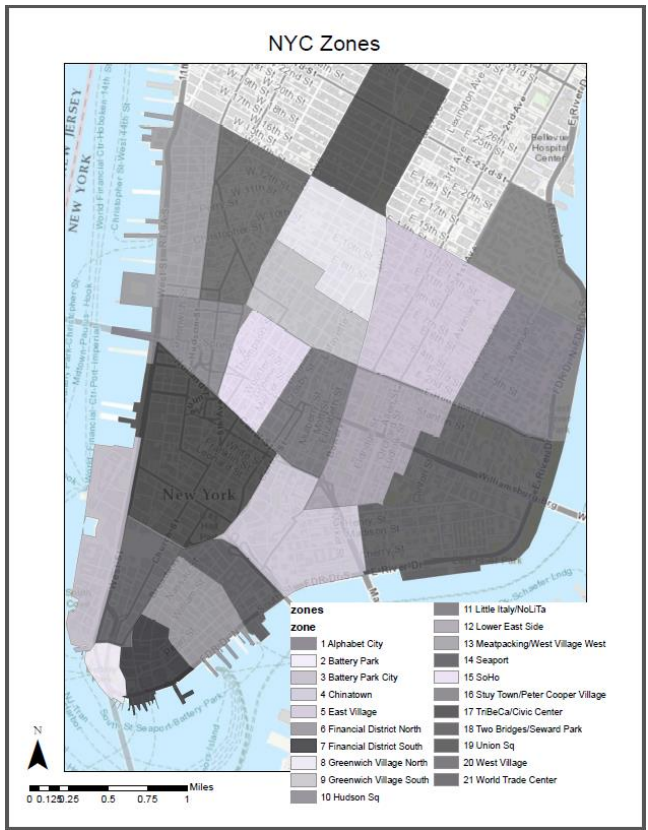

**Figure 6.** Study area of NYC taxi case study.

Within these 21 zones, 755 taxi trips were conducted during the study period. A set of potential routes are generated: single rider pickups and drop-offs as well as all possible combinations of two rider pickups and drop-offs. For each pair combination, the best route is generated. For example, two users $f$ and $g$ have origins and destinations $(O_f, D_f)$ and $(O_g, D_g)$ respectively. The shortest travel route from the following set is added to the route set: $\{(O_f - O_g - D_f - D_g), (O_f - O_g - D_g - D_f), (O_g - O_f - D_f - D_g), (O_g - O_f - D_g - D_f)\}$. For each pair of users we have three routes. If $|S|$ is the number of users, the number of routes in this case study is $|R| = \binom{|S|}{2} \times \frac{3}{2}$.

We break the study period into multiple time intervals $\Delta T$ and solve the problem as a static, multi-period assignment. In each interval, all the generated trips are pooled together and all the possible routes are generated to run the assignment model of Eq. (5) - (9). A $\Delta T$ is chosen to ensure that the static assignment problems within each interval contains non-empty stable outcomes in the base case. We initially assume $\Delta T$ equals 1 minute and keep dividing intervals with empty sets by two until the space of stable outcomes is non-empty. The same inferred stable intervals are used across the alternatives so that our comparison of single ride with shared ride policies is consistent. We find that an average stable interval length across the population of 46 seconds satisfies this objective.

It is assumed that each person's utility is equal to the travel time that they had in their observed single taxi ride plus the amount of fare they paid (which is known from the data). This utility is a lower bound of actual utility that each person has from doing his trip. This is a conservative choice for utility since these trips have already happened and if the actual utility was less, the user wouldn't have made the trip in the first place. In the numerical example, it does matter for each user which route they choose. As a direct route, a single ride obviously will have less travel time (more payoff) than a shared ride that has some delay due to detour.

In the numerical experiment of NYC taxi, the routes are explicitly generated beforehand. Based on generated routes the payoff value is independent of the user match ($x_{sr}$). There is no congestion effect from taxi rides on the road travel times – when a person chooses to ride one way or another, it will not impose crowding cost on another route at all (especially since taxis are only a small fraction of total traffic in lower Manhattan). The travel times do consider congested background traffic since the travel time matrices are drawn from Google API under the presence of congestion. Not considering congestion for shared taxi (and other fleet-oriented services) is normal and have been done in other shared ride studies (for a similar numerical example) such as Santi et al. (2013) and Alonso-Mora et al. (2017).

We consider the line capacity of route links equal to three passengers ($w_r = 3, \forall\, r \in R$). Each taxi user is assumed to be a single passenger so $q_s = 1$. All the cost and gain values are converted to monetary values. The value of time for travelling (TVOT) is assumed to be $0.40/min. The value of waiting time (WVOT) is assumed to be more than traveling value of time and is equal to twice that amount, $0.80/min. These values are conservative estimates based on numbers reported in Balcombe et al. (2004). Operating costs of routes are assumed to be $0.90/mile (based on an estimate of annual cost of $36,000 (including fixed and variable costs) and average annual mileage of 40,000 miles).

### *4.3. Results*
*4.3.1 Comparison of total shared taxi rides versus single taxi ride*

All the calculations are performed with a desktop computer with core i7 @3.40 GHz processor and 8.0 GB RAM. All the codes are written in MATLAB 2016a. The full set of results for all 755 users will be uploaded to GitHub (https://github.com/BUILTNYU) upon publication of this study.

We run the assignment game model on this data for the shared ride policy with a run time of 35.53 seconds. The total mileage of taxis for serving 755 users allowing shared rides is 1588.4 miles, compared to a total mileage for single-ride taxi in the original data of 1996.9 miles. This shows a *21% decrease in vehicle miles traveled*.

Second, the assignment shows that 523 of 755 users *(69%) decide to rideshare*. This is similar but lower than the 80% estimate from Santi et al. (2014). Our value is more pessimistic because it requires both users and operators to have sufficient incentive to switch. In Figure 7, the matches before (single taxi ride) and after (shared taxi ride allowed) matching taxi riders are shown.

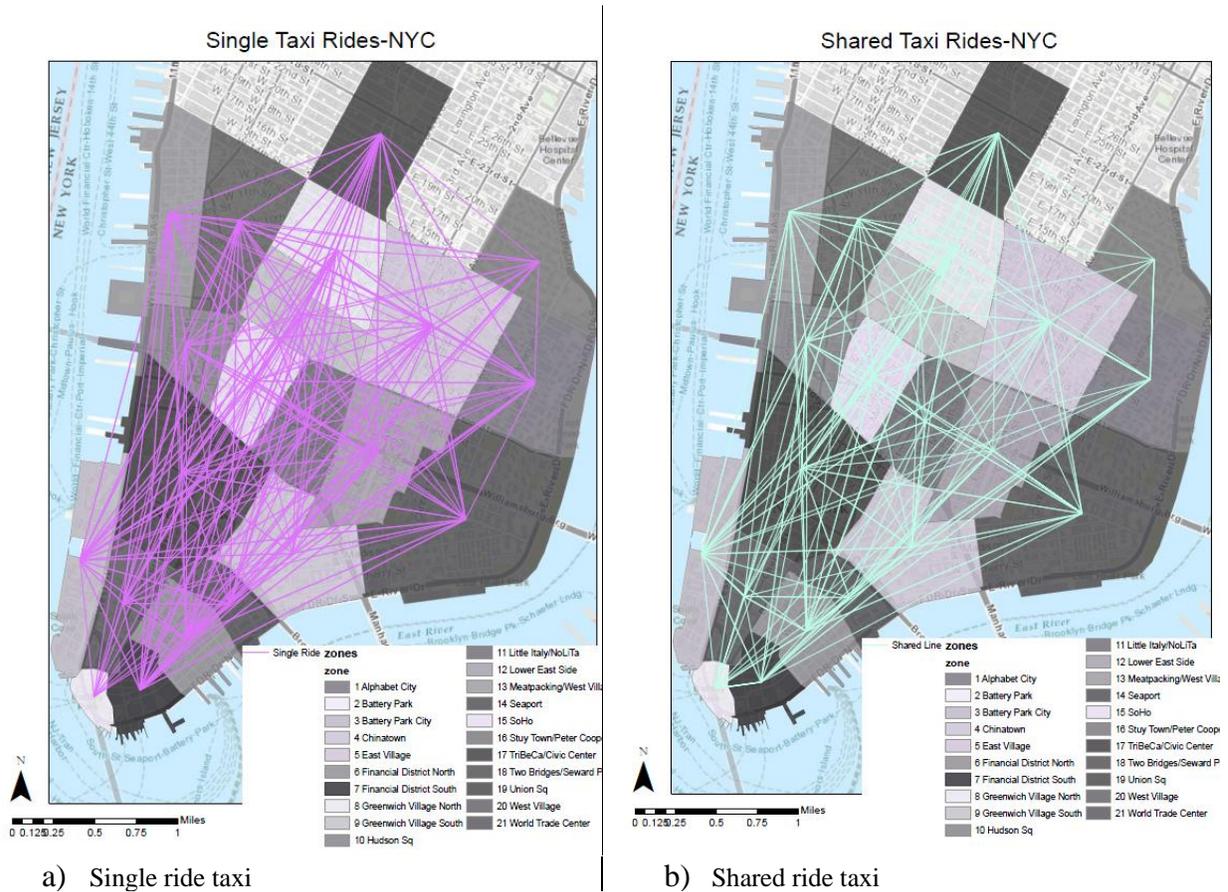

a)  Single ride taxi       b)  Shared ride taxi

**Figure 7**. NYC taxi assignment results for single and shared taxi riding.

In Figure 8, the generalized ticket price in three conditions of single riding taxi data, user-optimal shared taxi assignment, and operator-optimal shared taxi assignment are shown. The blue line represents the single ride taxi ticket price observed from the data. The red and green lines are ticket prices under operator- and user-optimal ticket pricing respectively (without minimum acceptable profit for operator and user). The space between these two lines is the space that a decision maker can use to evaluate stability of policies. With a positive minimum acceptable profit for users, the green and red lines would shift right, while a subsidy would shift them left.

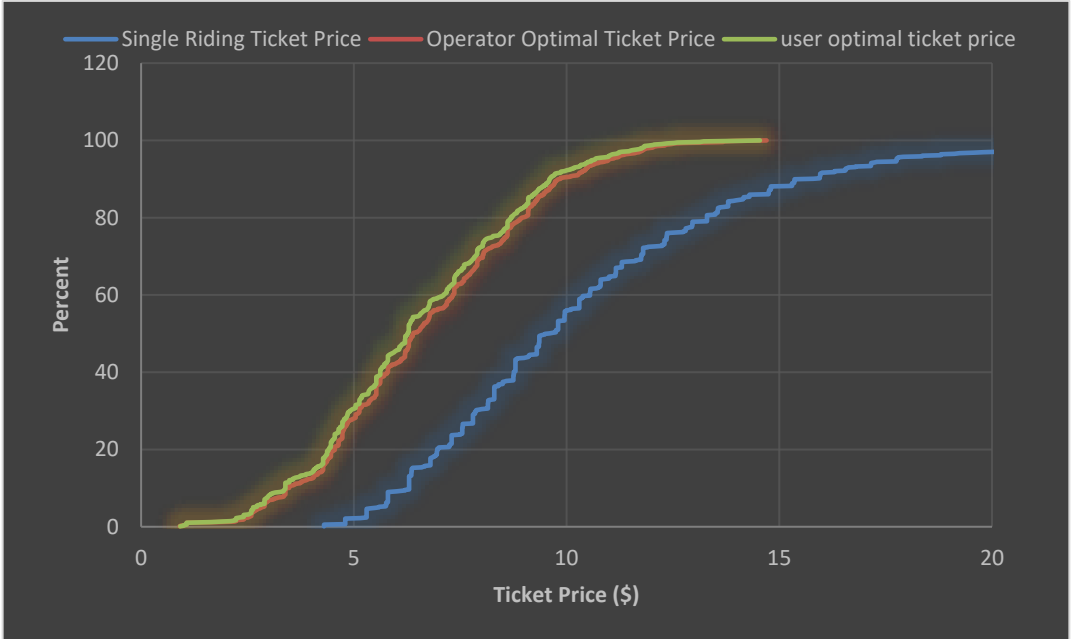

**Figure 8**. Ticket price percentage of users pay in three different scenario.

*4.3.2 Illustration of cost allocation mechanism design using the stable cost allocation space*

The benefit of the proposed model is that the set of stable outcomes is a convex polyhedron, which means that any weighted average price between the user-optimal and operator-optimal ticket price for each user and operator is also a stable price that can be used. In fact, this method provides a tool for explicitly pricing or setting design constraints for cost allocation mechanisms rather than blindly finding prices. This tool is a powerful method for pricing any matching situation in transportation systems, of which the NYC taxi is one example.

Consider the case of what happens if the shared ride taxi policy was incorporated with a fare price set at the user-optimal fare pricing level with the additional policy of requiring users to move up to a distance $D$ to be picked up. From the stable price space we know how much we can equivalently transfer operator cost to passenger cost without causing some users to break from the coalition. The total cost of the two minus the average operating cost savings per person should not exceed the operator optimal cost allocation.

This model allows us to determine a good threshold for $D$. If $D = 1\ block$, assuming 1 mile is approximately 20 blocks, pedestrian walking speed is 4 ft/s, and value of time of walking is approximately 1.5 of in-vehicle time (Balcombe et al., 2004), then the policy can impose up to an average of $0.44 on each traveler. Let us look at the distribution of gaps between the operator- and user-optimal pricing in Figure 9. It shows if the operator does not reallocate any costs to users along with the policy, there are only 99 out of 755 travelers (13.1%) who would have enough gap to stably absorb the additional walking cost without having to reduce operating cost to compensate. For the remaining 86.9%, however, the average gap is $0.03. If we want to ensure that no passenger is turned away by the walking policy, $(0.44 - 0.03) \times \frac{656}{755} = \$0.35$ per user should be transferred to the users from the cost savings from rerouting the vehicles. This conclusion illustrates how the operator can define performance benchmarks to evaluate a routing algorithm that requires the 1-block walking policy: e.g. one that saves less than $0.35 per user may not be stable without losing some users. The insight illustrates the strength of this modeling framework: for any general

transportation system with routes and line capacities, generalized cost allocation trade-offs between users and operators can be quantified and benchmarked for performance evaluation.

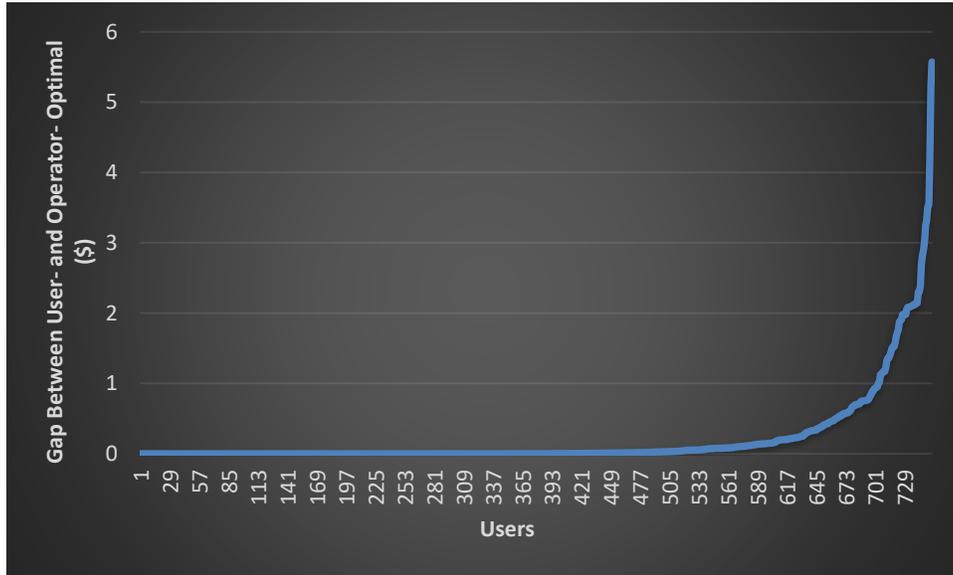

**Figure 9**. Sorted gap between user- and operator-optimal pricing under shared taxi policy.

*4.3.3 Detailed breakdown for select trips*

Lastly, we look more closely at what happens to single users. As shown in Figure 10, three users with OD pairs (11,13), (17,20) and (20,13) originally conducted their trips as solo taxi rides based on the data. Because of the availability of shared rides, they are now incentivized to share their rides together with the route ($\mathbf{17 \to 11 \to 20 \to 13}$). For each of these users there are other available options such as sharing their ride with other users or even riding alone.

The results for pricing users (11,13), (17,20) and (20,13) are shown in Table 7. In the single taxi riding condition, users (11,13), (17,20) and (20,13) are observed to pay $11.76, $6.30 and $8.15 respectively for their trips. The total profit of the operators from running these three trips is $8.34 + $3.6 + $7.34 = $19.28.

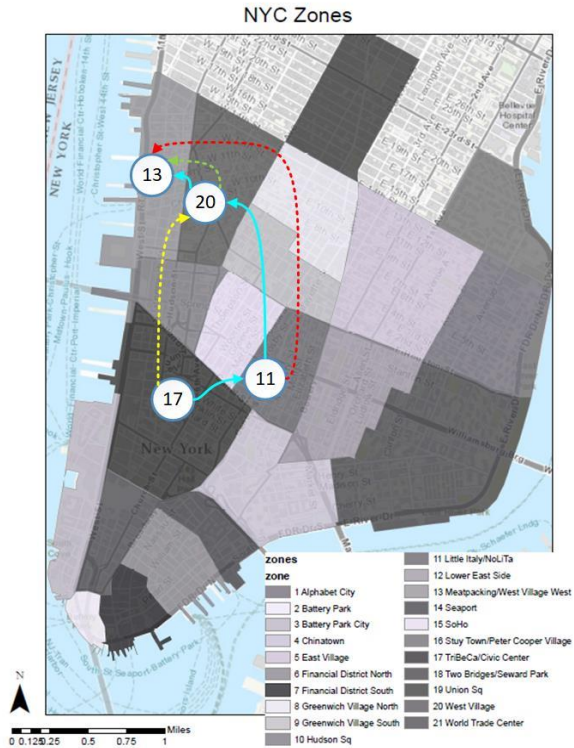

**Figure 10.** Illustration of single taxi riding (dashed) versus ridesharing for two select users (solid).

**Table 7.** Pricing of users (1,2) and (1,7) in single and ridesharing taxi riding

|  | Path | Results |  | Users | | |
|---|---|---|---|---|---|---|
|  |  |  |  | (11,13) | (17,20) | (20,13) |
| Single taxi riding | $1 \to 6$ | Travel time (min) |  | 15 |  |  |
|  |  | Ticket price ($) |  | 11.76 |  |  |
|  |  | User profit ($) |  | 0 |  |  |
|  |  | Operator profit ($) (ticket price − operation cost) |  | 8.34 |  |  |
|  | $5 \to 1$ | Travel time (min) |  |  | 13 |  |
|  |  | Ticket price ($) |  |  | 6.3 |  |
|  |  | User profit ($) |  |  | 0 |  |
|  |  | Operator profit ($) (ticket price − operation cost) |  |  | 3.6 |  |
|  | $5 \to 6$ | Travel time (min) |  |  |  | 6 |
|  |  | Ticket price ($) |  |  |  | 8.15 |
|  |  | User profit ($) |  |  |  | 0 |
|  |  | Operator profit ($) (ticket price − operation cost) |  |  |  | 7.34 |
| Ridesharing taxi riding | $1 \to 2 \to 7$ | Travel time (min) |  | 30 | 19 | 6 |
|  |  | Ticket price ($) | User optimal | $b + 9.34$ | $b + 7.82$ | $b + 3.13$ |
|  |  |  | Operator optimal | $b + 9.35$ | $b + 7.9$ | $b + 3.15$ |
|  |  | User profit ($) | User optimal | $8.41 - b$ | $3.67 - b$ | $7.39 - b$ |
|  |  |  | Operator optimal | $8.40 - b$ | $3.60 - b$ | $7.41 - b$ |
|  |  | Operator profit ($) (ticket price − operation cost) | User optimal | $3b$ | | |
|  |  |  | Operator optimal | $0.1 + 3b$ | | |

In the ridesharing condition, one operator serves these three users by a single route of (**17 → 11 → 20 → 13**). Stability conditions from Section 3.2 lead to the values shown in the lower portion of Table 7. The value $b$ is assumed to be a constant minimum acceptable profit for these two operators ($b < \$3.6$) that can serve these two users. For example, assuming $b = \$3$, the operator profit from operating one route will vary from $9 in user-optimal ticket pricing to $9.1 in operator optimal ticket pricing. These values compare to $8.34, $3.6 and $7.34 for each route in single riding taxi.

Under the same assumption for $b$, user benefits for users (11,13), (17,20) and (20,13) would vary respectively from $9.34, $7.82 and $3.13 in operator-optimal ticket pricing to $9.35, $7.9 and $3.15 in user optimal ticket pricing condition. This value compares to the baseline of $0 in single taxi riding (since we set the utility to be the single ride cost). This quantification of the benefits can be used by public agencies to deploy pricing schemes to encourage people to shift to shared rides.

### *4.4. Computational performance*

Like other capacitated multicommodity flow problems, the performance of the mixed integer programming solution method depends on the number of routes generated for each OD pair. In the assignment model of Eq. (5) - (9), for a set of $n$ unique user ODs, we have $n$ constraints (6), $\frac{n(3n-1)}{2}$ constraints (7), and $\frac{n(n+1)}{2}$ constraints (8). So, for $n$ unique user OD pairs, we have constraints on the order of $n^2$. With an increase in the number of unique user OD pairs, the number of constraints increases drastically. As suggested in section 3.3, one way to address this computational challenge is to introduce implicit enumeration using column generation (Barnhart et al., 1994), which we will examine more closely in future research.

To provide a measure of the computational complexity of solving the assignment problem using mixed integer program without any decomposition, we tested the method on different numbers of unique user OD pairs. In a rectangular study area, we generated origin and destination locations for $n$ requests (in Euclidian space). In a similar process to section 4.2, for each OD pair, a set of potential routes are generated: single rider pickups and drop-offs as well as all possible combinations with other OD. For each pair combination, the best route is generated. For example, two OD pairs of $f$ and $g$ have origins and destinations $(O_f, D_f)$ and $(O_g, D_g)$ respectively. The shortest travel route from the following set is added to the route set: $\{(O_f - O_g - D_f - D_g), (O_f - O_g - D_g - D_f), (O_g - O_f - D_f - D_g), (O_g - O_f - D_g - D_f)\}$. Figure 11 shows how the solution time and number of constraints increase with number of OD pairs.

We can see that the solution time goes up exponentially while the number of constraints increases quadratically. This computational analysis confirms the need to explore decomposition algorithms in future research.

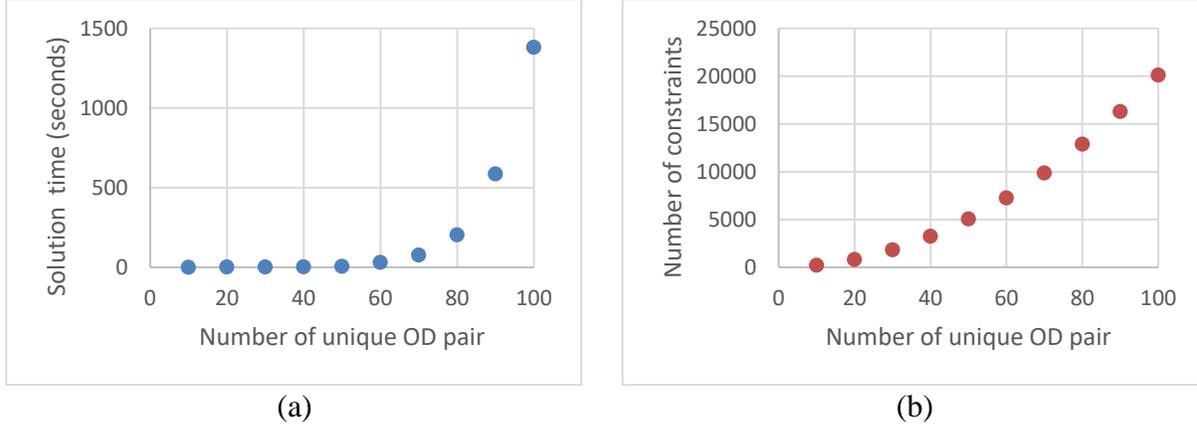

(a)                                                  (b)

**Figure 11**. (a) Solution time and (b) number of constraints of assignment problem for different number of unique OD pair.

## 5. Conclusion

There is currently no computational transportation assignment model for a broad class of mobility services in MaaS because existing models do not jointly model user and operator behavior. We propose a generalized assignment game model to overcome this problem, which has the potential to transform the transportation planning practice with a new quantitative tool for evaluating emerging mobility services in a smart cities era. We show that using this model has several benefits over the state of the art:

- It is generalized prescriptive assignment methodology that is not restricted to just one type of mobility service and can handle both decentralized and centralized operators with line capacity effects.
- Unlike most of the studies that assume a cost allocation policy or mechanism, the proposed model outputs a stable payoff space for post evaluation of cost allocation mechanisms. We show that decision-makers need to have either full control or partial control (through sequential design) of the assignment and cost allocation mechanisms to use the model for system design.
- By considering the behavior and incentives of both users and operators, the model allows policymakers and operators to evaluate the stability and pricing (in general costs: fare, access, wait, transfer, in-vehicle time) requirements of operating policies and cost allocation mechanisms.
- In the conventional assignment game, a user uses a whole product of a seller. In transportation, however, passengers may be matched only to sections of a route. This makes the stability condition of our proposed model markedly different from prior assignment games. In section 3.2 the stability principles for this proposed model are derived.

We examined our model through several different examples. The first set of examples deal with a 4-node fixed route transit network. The last set of examples investigates a case study calibrated from real NYC taxi data as a proof-of-concept to illustrate the proposed methodology. Using the proposed model, we assigned taxi users to share their cab with a range of stable pricing. In this study with 755 trips from downtown Manhattan, we showed that 69.3% (523 trips) of users are willing to share their taxis. Ridesharing of this 69.3% of users decreases the vehicle miles traveled by 20.5%.

Several directions for future research are possible. The model can be extended to consider routes having to share limited space with congestion effects. Generally, the focus of MaaS systems is on vehicle and line capacity, but certainly in highly congested cities with limited space and very large fleets it is important to study the congestion effects that different operators have on each other based on serving travelers. For example, in NYC there is consideration of a surcharge on taxis and for-hire vehicles because they crowd the streets which slow down bus routes.

A second issue to consider is route enumeration. The model assumes matching users to operator routes, which requires having operator routes beforehand. This is a problem in the line planning literature as well. Solutions include using heuristics (Ceder and Wilson, 1986) and column generation methods (Barnhart et al., 1994; Borndörfer et al., 2007) to generate routes. Future research may consider endogenous equilibration of the route set as well (e.g. as studied by Watling et al., 2015, and Rasmussen et al., 2015).

A third potential future study is dynamic assignment. The current methodology and examples deal with static assignment games, even though dynamic assignment would be more appropriate for some analysis. We know that several ridesharing companies currently need dynamic assignment since their demand is not known in advance. As a result, an interesting direction for future studies can be also considering stable matching and its dynamic pricing that considers sequential cost allocations (e.g. Gopalakrishnan et al., 2016).

The model can be used to evaluate flexible route decisions for multimodal transit services. Instead of matching a user's OD pair to a portion of an operator's route, we can match portions of operator routes to portions of a user's route, allowing for truly multimodal assignment. In those cases, the operators may have to cooperate by considering cost transfers with each other (e.g. fare bundles) to serve the passengers. This becomes a many-to-many assignment game like in Sotomayor (1999), but with line capacity effects on top of that.

With the increasing devastation of catastrophic disasters, it has become necessary to be better prepared for these kinds of events. For future studies, we will also consider stochastic scenarios for risk pooling (cost allocations between operators) in the case of such events.

## Acknowledgments

This research was supported by a grant from the National Science Foundation, CMMI-1634973. We thank the four anonymous referees of this journal who contributed comments to improving the quality of this manuscript.

## Appendix. Model variations

Variations can be modeled. For example, direct services like taxis may be represented by single link routes with some dynamic filtering of available routes. A route can also be replaced with a cycle without altering the model. If the operator is a single centralized decision-maker, then the matching would be made between the set of users with a single seller. Operators may be toll road operators, public transport providers, taxis, or TNCs, among others mentioned earlier. The following sections describe how centralized decision-making and vehicle-route framework can be modeled.

*A.1. Operator-route case*

Although the base model assumes each operator is a separate route, it is also possible to consider operators as collections of routes, or even as a system-wide centralized agency. This section demonstrates how that can be accomplished using our underlying model.

In the centralized case, as before, set $P$ represents the set of operators. Operator $p \in P$ owns the set of routes to operate ($R_p$). $O(r)$ is the operator that owns route $r$ (e.g. $O(r \in R_p) = p$). In a decentralized situation, each operator owns only one route. The main assignment model (5) - (9) remains valid for the centralized case because users and routes would be matched to the routes that generate maximum payoff together. The key difference between the centralized and decentralized case is in their stability definitions.

A modification is made to the stability defined by the convex area in Eq. (29) to (33). In a centralized case, the stability related to different operators (as opposed to different routes) should be checked. While an operator wants to make sure that their users won't switch to another operator's route, they don't mind if their user switches to a different route belonging to the same operator. This is a looser stability condition which can be obtained by replacing Eq. (29) with Eq. (34). The set of stable outcomes in this case includes the stable outcomes that are defined in decentralized case.

$$\sum_{s \in G_r} u_s + v_r \geq \sum_{s \in G_r} a_{sr} - C_r \qquad \begin{array}{l} \forall G_r \text{ and } r \in R \\ \exists s \in G_r : O(r) \neq O(C(s,x)) \end{array} \qquad (34)$$

In Eq. (34), stability is checked to ensure a user cannot generate more payoff with a route of another operator. This change ensures the stability over different operators. This variant can be used to model either centralized operators with multiple routes or exogenous coalitions of operators. Handling endogenous coalition formation between operators would require extending this model to include network flow games, which we reserve for future research.

*A.2. Operator-vehicle-route case*

We introduce another variation of the model under the operator-route case. We model a market where operators own a fleet of vehicles. Set $P$ represents the set of operators. Operator $p \in P$ owns a set of vehicles ($V_{p \in P}$) with which to operate from a set of candidate paths. For example, a train can be operated along certain paths and a taxi vehicle can be operated along different paths at different times. In this case the two sides of the market are users and vehicle-paths. Let's call $r_{pi}$ the set of candidate paths for vehicle $i \in V_p$. $R_p$ is the set of candidate paths for all the fleet of operator $p$ ($\bigcup_{i \in V_p} r_{pi} = R_p$). The assignment model of Eq. (5) - (9) remains the same for this vehicle based system except it needs one more set of constraints (Eq. (35)). For each $r_{pi}$ set, a maximum of one (non-k) path is chosen (i.e. each vehicle cannot have more than one $r$ where $x_{kr} = 0$ because that implies more than one path operating at the same time per vehicle).

$$\sum_{r \in r_{pi}} (1 - x_{kr}) \leq 1 \qquad \forall i \in V_p, \forall p \in P \qquad (35)$$

Another modification that should be made for this case is related to the stability conditions. Since the introduced case of this section is a centralized decision making system, the stability equations of Eq. (34) should be used instead of Eq. (29).

*A.3. Example of operator-vehicle-path case*

A 3-node network is considered. The network is identical to the 4-node network of Section 3.4.1 but without the 4[th] node. The link costs are in units of operating miles. There are two operators, each with a fleet of 2 vehicles, and each vehicle has a passenger capacity of 2. The initial locations of the fleets are node 1 for operator 1 and node 3 for operator 2. The vehicle speeds are $40\ mph$. The set of users' OD is $S = \{(1,2), (1,3), (2,3), (3,2)\}$. The utility of all passengers is set to the value of \$20 ($U_{sr} = 20\$ \ \forall s \in S\setminus\{k\}, r \in R$). The value of time is \$0.4/min and cost of operation to \$0.9/mile. The direct cost that each passenger experiences is the waiting time plus travel time. Waiting time is multiplied by 1.25. Operator 1's vehicles each have 6 candidate paths $r_{1i \in V_1} = \{(1 \to 2), (1 \to 3), (1 \to 2 \to 3), (1 \to 3 \to 2), (1 \to 2 \to 3 \to 1), (1 \to 3 \to 2 \to 1)\}$. Operator 2's vehicles have 5 candidate paths $r_{1i \in V_2} = \{(3 \to 2), (3 \to 1 \to 2), (3 \to 2 \to 1), (3 \to 1 \to 2 \to 3), (3 \to 2 \to 1 \to 3)\}$. The results of assignment and ticket prices is shown in the Table A1.

**Table A1.** Results of assignment and payoff allocation in centralized case

|  | Chosen path | User (O,D) | | | |
|---|---|---|---|---|---|
|  |  | (1,2) | (1,3) | (2,3) | (3,2) |
| Vehicle 1 of operator 1 | $1-2-3$ | 1 | 1 | 1 |  |
| Vehicle 1 of operator 2 | $3-2$ |  |  |  | 1 |
| Waiting time (min) |  | 0 | 0 | 4.5 | 0 |
| Travel time (min) |  | 4.5 | 7.5 | 3 | 3 |
| User optimal ticket price |  | $1.5 + b_1$ | $1.5 + b_1$ | $1.5 + b_1$ | $1.8 + b_1$ |
| Operator optimal ticket price |  | $5.55 + b_2 - b_1$ | $2.25 + b_2 - b_1$ | $8.55 + b_2 - b_1$ | $1.8 + b_1 - b_2$ |

Based on Table A1, Vehicle 1 of Operator 1 is assigned to path $1 \to 2 \to 3$ to serve passengers (1,2), (1,3) and (2,3). Vehicle 1 of Operator 2 is assigned to path $3 \to 2$ to serve the passenger (3,2). $b_1$ and $b_2$ are the minimum acceptable profits for Operators 1 and 2. The operator-optimal ticket price for a passenger is a function of the minimum acceptable profit by the matched operators. For a smaller value of rival operator acceptable profit, the operator would charge their passenger less to keep the coalition stable.

Minimum acceptable profit is a policy making tool for operators to compete in a market. For example, sometimes the operator loses some gains over some routes to compete on other routes in different timetables. A newcomer operator to a market may want to penetrate the market by maximizing the number of matches initially without concern for profit. In this context, even negative values for minimum acceptable profit have economic interpretation.